\documentclass[jpcm,showpacs,twocolumn]{revtex4}
\usepackage{amsfonts}
\usepackage{amsmath}
\usepackage{amssymb}
\usepackage{graphicx}

\begin{document}

\title{Dephasing of electrons in the Aharonov-Bohm interferometer with a single-molecular vibrational junction}
\author{Wenxi Lai, Yunhui Xing, and Zhongshui Ma}
\affiliation{School of Physics, Peking University, Beijing 100871, China}

\begin{abstract}
\begin{center}
\textbf{Abstract}
\end{center}
Phase relaxation of electrons transferring through an electromechanical transistor is studied using the Aharonov-Bohm interferometer. With the approach of quantum master equation, the phase properties of an electron are numerically analyzed based on the interference fringes. Coherence of electron is partially destroyed by its scattering on excited levels of the local nanomechanical oscillator. Transmission amplitudes with respect to two adjacent mechanical vibrational levels have a phase difference of $\pi$. The character of phase shift by $\pi$ depends on the oscillator frequency only and is robust for the wide range variance of the applied voltage, tunneling length and damping rate of the mechanical oscillator.
\end{abstract}

\maketitle

\begin{center}
\textbf{1. Introduction}
\end{center}

Discrete quantum states and coherent electronic transport are two important
properties of mesoscopic conductors. In a recent experiment of a single-molecule $C_{60}$
transistor, current steps because of the quantum behavior of nanomechanical motion was observed~\cite{Park}.
In the sequent years, the electron transport through the mechanical vibrational junctions becomes
a subject of much topical interests. Such systems were fabricated with a gold nanoparticle
oscillator quite recently~\cite{Moskalenko}. More fascinating features of the electromechanical
systems have been predicted with theoretical approaches, such as negative differential
conductance~\cite{Boese, McCarthy}, shuttling effect~\cite{Gorelik:1998,Novotny:2003},
super and sub-Poissonian Fano factor~\cite{Novotny:2004,Koch,Haupt,Lai},
and spintronic transport~\cite{Fedorets,Twamley,Wang}, etc.

By applying the Aharonov-Bohm (AB) interferometer with a quantum dot (QD) embedded in one arm,
coherence of electron tunneling through the QD is studied in experiments~\cite{Yacoby,Schuster,Cernicchiaro}.
These experiments show that a fixed QD supports coherent transport and causes a phase shift to an electron. When a QD is allowed to mechanically oscillate around its equilibrium point, an electron transferring through the dot would be
accompanied by random absorption or emission of phonons. Phase property in the mechanical vibration assisted electron tunneling is still an open interesting question. The vibrational motion in the system can be modeled to a good approximation as
a monochromatic oscillator. It is different from thermal fluctuating bosonic
baths which cause decoherence to local electronic states of charge~\cite{Stavrou,Grodecka-Grad} and spin~\cite{Roszak,Hu}. As recently reported, a single vibrational mode of QD-array enhances electron transport and partially preserves its phase
information~\cite{Milburn}. It is worth mentioning that coherent transport of electrons in QDs
is also sensitive to spin flip, electron-electron interaction and external detectors~\cite{Aleiner,Buks,Sprinzak,Konig:2001,Konig:2002,Aikawa,Khym,Moldoveanu,Rohrlich}.

In this work, we shall investigate dephasing of electrons induced by the electromechanical vibration in the single-molecular transistor. It is implemented by embedding a harmonically movable QD in one (target) arm of the AB interferometer and locating a fixed QD in the other (reference) arm. The reason of using the QD in the reference arm is that phase shift corresponding to each discrete level of the mechanical oscillator can be observed by changing the gate voltage of the reference QD. The previous research close to our issue of interest is the which-path detector of charge by using a cantilever~\cite{Armour:2001,Armour:2002}. This detector is based on dot-cantilever coupling which causes remarkable dephasing to the electrons. In their model, the dot-lead coupling does not depend on oscillator position. Whereas, the position dependence of the coupling is considered in our system since it is significant for the electromechanical shuttle junction~\cite{Moskalenko,Gorelik:1998}. In our previous report, we derived a fully quantum mechanical master equation to describe the electromechanical system~\cite{Lai}. In the equation we considered both diagonal and off-diagonal density matrix elements for the states of vibrational levels, and revealed that the off-diagonal terms have important contribution to the electronic current. Here, we shall develop this approach to describe the influence of the electromechanical system to the AB interference. To obtain knowledge of the phase properties of electron scattering on the vibrational junction, interference fringe as a function of an external magnetic flux through the AB ring will be analyzed with respect to various parameters. In the following, this article will discuss that coherence of electron in the AB ring is suppressed due to excitation of the vibrational mode in the transport process. The suppression becomes more serious when one increases the bias voltage or decreases the oscillator damping rate. It is shown that by moving the energy level of the reference QD, global phase shift in the transmission amplitude can be observed from the change of interference fringe. The transmission amplitudes corresponding to two neighboring resonant levels of the molecular junction are off-phase by $\pi$. This phase difference causes destructive interference to propagating waves and destroys the coherence of electron.

This paper is organized as follows: In section 2, we derive the master equation for the model of the AB interferometer with a single-molecular transistor in one arm. In section 3, the method of our numerical calculation is introduced. In section 4, we present solutions for various parameters. Influence of the electromechanical system to the interference fringe is analyzed. In addition, the phase shifts of electronic propagating waves through the resonating QD will be illustrated. Finally, conclusions are given in section 5.

\begin{center}
\textbf{2. Model and equation of motion}
\end{center}

\begin{figure}[!htb]
\centering \includegraphics[width=8cm]{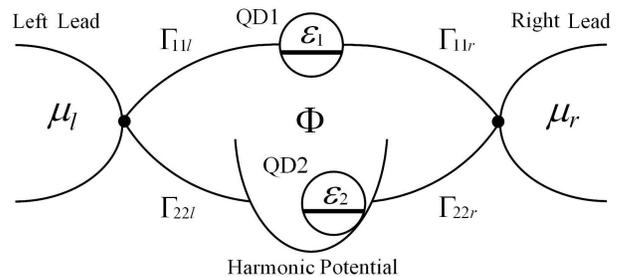}
\caption{Two single-level quantum dots connect to two leads in parallel, which enclose a magnetic flux $\Phi$ for Aharonov-Bohm interference. The upper dot is fixed and the lower dot is bounded by an harmonic potential and movable in the horizontal direction.}
\label{sys}
\end{figure}

The schematic structure of our AB interferometer is illustrated in figure~\ref{sys}. It contains two single-level QDs coupled to two electronic leads in parallel. One of them with energy $\varepsilon _{1}$ (QD1) is fixed in the upper arm and the other with energy $\varepsilon _{2}$ (QD2) is localized in the lower arm. The two arms and the electrodes enclose a magnetic flux $\Phi $ passing through the loop-plane. The QD2 is assumed to be bounded in a harmonic potential, which consists the electromechanical shuttle junction. The QD1 in the upper arm provides a reference path. We consider both inter and intra-dot Coulomb blockade limits in order to make sure that electrons propagate through the two-path interferometer one by one. Spin degree of freedom is not involved in our approach. The Hamiltonian can be written in the form of
\begin{eqnarray}
H=H_{leads}+H_{dots}+H_{mech}+H_{tun},
\end{eqnarray}
where
\begin{eqnarray}
H_{leads}=\sum_{k;y=l,r}\xi _{yk}d_{yk}^{\dag }d_{yk}
\end{eqnarray}
describes the noninteracting electrons in the left ($y=l$) and right (y=r) leads. $d_{yk}^{\dag }$ and $d_{yk}$ are creation and annihilation operators of electrons with momentum $k$ and energy $\xi _{yk}$.
In the Hamiltonian
\begin{eqnarray}
H_{dots}=\epsilon _{1}c_{1}^{\dag }c_{1}+(\epsilon _{2}-\hbar \Omega
(a^{\dag }+a))c_{2}^{\dag }c_{2},
\end{eqnarray}
$c_{i}^{\dag }$($c_{i}$) is the creation (annihilation) operator of QD$i$ ($i$=1,2). Here, the last term is the work taken by charged QD2 when it moves a distance $x_{0}(a^{\dag }+a)$ in an external electric field $V/d$. The coupling coefficient is given by $\Omega=eVx_{0}/\hbar d$ with the bias voltage $V$ and effective distance $d$ between the two electrodes. $e$ is the absolute value of the electron charge and $x_{0}$ is the zero point position uncertainty $\sqrt{\hbar/2m\omega_{0}}$ of the oscillator with frequency $\omega_{0}$ and effective mass $m$. The nanomechanical vibration is treated in the quantum regime as
\begin{eqnarray}
H_{mech}=\hbar \omega _{0}a^{\dag }a+\sum_{k}\hbar \omega _{k}b_{k}^{\dag
}b_{k}+\sum_{k}\hbar (gb_{k}^{\dag }a+h.c.).
\end{eqnarray}
$a$ ($a^{\dag }$) and $b_{k}$ ($b_{k}^{\dag }$) are annihilation (creation) operators for the vibrational mode and its thermal bath, respectively. $\omega _{k}$ denotes frequency of mode $k$ in the thermal bath which is coupled to the oscillator with a coefficient $g$.
Tunneling through the two QDs is represented by
\begin{eqnarray}
H_{tun} &=&\hbar \sum_{k;y=l.r}(T_{2y}e^{-iS_{y}\phi /4}e^{S_{y}\alpha
(a^{\dag }+a)}d_{yk}^{\dag }c_{2}+h.c.)  \notag \\
&&+\hbar \sum_{k;y=l,r}(T_{1y}e^{iS_{y}\phi /4}d_{yk}^{\dag }c_{1}+h.c.).
\end{eqnarray}%
The tunneling amplitudes between the two leads and QD1 are given by $T_{1y}e^{iS_{y}\phi /4}$ ($S_{l\left(r\right) }=-1\left( +1\right)$) and its complex conjugate, where the phase $\phi$ is related to the magnetic flux $\phi=2\pi \Phi /\Phi _{0}$ with the flux quantum $\Phi _{0}=h/e$. The tunneling amplitude with respect to QD2 is written as $T_{2y}e^{-iS_{y}\phi /4}e^{S_{y}\alpha(a^{\dag }+a)}$ which exponentially depends on the position of the oscillator. The parameter $\alpha$ is inverse ratio of the tunneling length $\lambda$ as $\alpha=x_{0}/ \lambda$.

We describe the interference process using the quantum master equation. It is extended from the equation given in our earlier publication where a more detail derivation can be found~\cite{Lai}. The state of total configuration is described with the density matrix $\rho _{T}(t)$ which satisfies the Liouville-von Neumann equation
\begin{eqnarray}
i\hbar \dot{\rho}_{T}(t)=[H(t),\rho _{T}(t)].
\end{eqnarray}
Both of the electronic leads and the thermal bath are assumed to be in equilibrium all the time and descried by the time independent equilibrium density matrices $\rho _{L}$ and $\rho _{B},$ respectively. Assuming the initial state as $\rho _{T}(0)=\rho (0)\rho _{L}\rho _{B}$, we can write the state at time $t$ in the form $\rho _{T}(t)=\rho (t)\rho _{L}\rho _{B}$ under the Born approximation. $\rho (t)$ is the reduced density matrix of the system which consists of the two QDs and the mechanical oscillator. Iterating equation (6) in the interaction picture to the second order and performing trace over the leads $(Tr_{L})$ and the bath $(Tr_{B})$ variables, we obtain the master equation for the reduced density matrix of the system in the Markov approximation as
\begin{eqnarray}
\dot{\rho}^{v}=\varrho _{0}^{v}+\varrho _{1}^{v}+\varrho_{2}^{v}+\varrho _{12}^{v}+\varrho _{d}^{v}.
\end{eqnarray}
On the right hand side of equation (7), $\varrho_{0}^{v}$ denotes evolution term of the system in which QD2 is coupled to the harmonic oscillator. $\varrho _{1}^{v}$ describes the contribution from direct tunneling by QD1 in the absence of QD2. $\varrho _{2}^{v}$ is just the right hand side of the master equation in our previous work~\cite{Lai}, representing the contribution from vibration assisted transfer through QD2 alone. $\varrho _{12}^{v}$ is coherent term of the transport involving the two dots and $\varrho_{d}^{v}$ accounts for dissipation of the vibrational mode. Explicit expressions of these terms read
\begin{widetext}
\begin{eqnarray}
\varrho _{0}^{v}=\frac{1}{i\hbar }[\varepsilon _{1}c_{1}^{\dag
}c_{1}+\hbar \omega _{0}a^{\dag }a+(\varepsilon _{2}-\hbar \Omega (a^{\dag }+a))c_{2}^{\dag
}c_{2},\rho ^{v}] ,
\end{eqnarray}
\begin{eqnarray}
\varrho _{1}^{v} &=&\frac{1}{2}\sum_{y=l,r}[\Sigma^{in}_{11y}(0,0,1)(2c_{1}^{\dag }\rho ^{v^{+}_{y}}c_{1}-c_{1}c_{1}^{\dag }\rho ^{v}-\rho^{v}c_{1}c_{1}^{\dag})
+\Sigma^{out}_{11y}(0,0,1)(2c_{1}\rho ^{v^{-}_{y}}c_{1}^{\dag }-c_{1}^{\dag }c_{1}\rho ^{v}-\rho
^{v}c_{1}^{\dag}c_{1})],
\end{eqnarray}
\begin{eqnarray}
\varrho_{2}^{v}&=&\frac{1}{2}e^{\alpha ^{2}}\sum_{y=l,r}\sum^{\infty}_{m_{1},n_{1},m_{2},n_{2}=0}\frac{(S_{y}\alpha
)^{m_{1}+n_{1}+m_{2}+n_{2}}}{m_{1}!n_{1}!m_{2}!n_{2}!} \notag \\
&&\times [\Sigma^{in}_{22y}(m_{1},n_{1},2)(A_{m_{2}n_{2}}^{+}\rho ^{v_{y}^{+}}A_{n_{1}m_{1}}^{-}+A_{m_{1}n_{1}}^{+}\rho ^{v_{y}^{+}}A_{m_{2}n_{2}}^{-}-A_{m_{2}n_{2}}^{-}A_{m_{1}n_{1}}^{+}\rho
^{v}-\rho^{v}A_{n_{1}m_{1}}^{-}A_{m_{2}n_{2}}^{+}) \notag \\
&&+\Sigma^{out}_{22y}(m_{1},n_{1},2)(A_{m_{2}n_{2}}^{-}\rho ^{v_{y}^{-}}A_{m_{1}n_{1}}^{+}+A_{n_{1}m_{1}}^{-}\rho^{v_{y}^{-}}A_{m_{2}n_{2}}^{+}-A_{m_{2}n_{2}}^{+}A_{n_{1}m_{1}}^{-}\rho^{v}
-\rho^{v}A_{m_{1}n_{1}}^{+}A_{m_{2}n_{2}}^{-})],
\end{eqnarray}
\begin{eqnarray}
\varrho _{12}^{v}& =&\frac{1}{2}e^{\alpha ^{2}/2}\sum_{y=l,r}e^{iS_{y}\phi /2}\sum^{\infty}_{m,n=0}\frac{(S_{y}\alpha
)^{m+n}}{m!n!} \notag \\
&&\times[\Sigma^{in}_{12y}(0,0,1)(A_{mn}^{+}\rho^{v_{y}^{+}}c_{1}-\rho ^{v}c_{1}A_{mn}^{+})
+\Sigma^{out}_{12y}(0,0,1)(c_{1}\rho^{v_{y}^{-}}A_{mn}^{+}-A_{mn}^{+}c_{1}\rho^{v}) \notag\\
&&+\Sigma^{in}_{12y}(m,n,2)(A_{mn}^{+}\rho^{v_{y}^{+}}c_{1}-A_{mn}^{+}c_{1}\rho^{v})
+\Sigma^{out}_{12y}(m,n,2)(c_{1}\rho^{v_{y}^{-}}A_{mn}^{+}-\rho ^{v}c_{1}A_{mn}^{+})]+h.c.,
\end{eqnarray}
\end{widetext}
and
\begin{equation}
\varrho _{d}^{v}=\Xi^{in} D[a^{\dag }]\rho ^{v}+\Xi^{out} D[a]\rho ^{v}.
\end{equation}
The degree of freedoms in the electronic leads and the thermal bath are assumed to be continuous with densities of states $N_{y}(\xi_{yk})$ and $D(\omega_{k})$, respectively. In the above equations, the coefficients corresponding to particles hop into or out of the system are composed of the integrals over these reservoir variables via
\begin{eqnarray}
\Sigma^{in}_{ijy}(m_{1},n_{1},z) &=&\int d\xi_{yk}\Gamma_{ijy}(\xi_{yk})f_{y}(\xi_{yk}) \notag \\
&&\times\delta(\xi_{yk}-\epsilon_{z}-(m_{1}-n_{1})\hbar\omega_{0}), \notag
\end{eqnarray}
\begin{eqnarray}
\Sigma^{out}_{ijy}(m_{1},n_{1},z) &=&\int d\xi_{yk}\Gamma_{ijy}(\xi_{yk})(1-f_{y}(\xi_{yk})) \notag \\
&&\times\delta(\xi_{yk}-\epsilon_{z}-(m_{1}-n_{1})\hbar\omega_{0}), \notag
\end{eqnarray}
\begin{equation*}
\Xi^{in}=\int d\omega_{k} \gamma(\omega_{k})n_{B}(\omega_{k})\delta(\omega_{k}-\omega_{0}),
\end{equation*}
and
\begin{equation*}
\Xi^{out}=\int d\omega_{k} \gamma(\omega_{k})(1+n_{B}(\omega_{k}))\delta(\omega_{k}-\omega_{0}).
\end{equation*}
Here, $\Gamma_{ijy}(\xi_{yk})=2\pi N_{y}(\xi_{yk})T_{iy}^{\ast }T_{jy}$, $\gamma(\omega_{k})=2\pi D(\omega_{k})|g|^{2}$ and $i,j=1,2$. We have the Fermi-Dirac distribution function in lead $y$, $f_{y}(\xi_{yk})=[e^{(\xi_{yk}-\mu _{y})/k_{B}T}+1]^{-1}$ and the Bose-Einstein distribution function of the thermal bath, $n_{B}(\omega_{k})=\left( e^{\hbar \omega_{k}/k_{B}T}-1\right) ^{-1}$, where $T$ is temperature and $k_{B}$ is the Boltzmann constant. In the above equation we have defined $A_{mn}^{-}=c_{2}(a^{\dag })^{m}(a)^{n}$, $A_{mn}^{+}=c_{2}^{\dag}(a^{\dag })^{m}(a)^{n}$, where $A_{mn}^{+}$ $(A_{mn}^{-})$ describes that an electron hops into (out of) QD2 accompanied by creation of $m$ phonons and annihilation of $n$ phonons. $v$, $v_{y}^{+}=v+(1+S_{y})/2$ and $v_{y}^{-}=v-(1+S_{y})/2$ indicate the number of electrons accumulated in the right lead. They are achieved by the following way: $Tr_{L}(d^{\dag}_{rk}d_{rk}\rho_{L})$ and  $Tr_{L}(d_{rk}\rho_{L}d^{\dag}_{rk})$ contain different information about the number of electrons in the right lead when the number is not infinite. Assuming $v$ electrons are in the right lead, then the number of electrons in this lead can be expressed by $\rho^{v}f_{y}(\xi_{rk})=\rho Tr_{L}(d^{\dag}_{rk}d_{rk}\rho_{L})$ and $\rho^{v+1}f_{y}(\xi_{rk})=\rho Tr_{L}(d_{rk}\rho_{L}d^{\dag}_{rk})$. In the same way, we have $\rho^{v-1}f_{y}(\xi_{rk})=\rho Tr_{L}(d^{\dag}_{rk}\rho_{L}d_{rk})$. The density matrix satisfies $\sum_{v=0}^{\infty}\rho^{v}=\rho$. The above method is equivalent to the counting approach in the many-body Schrodinger equation~\cite{Gurvitz:1996}, representing how many particles arrive at the collector. $D[a]\rho ^{v}$ is a super operator acting on the density matrix $\rho ^{v}$ by $D[a]\rho^{v}=a\rho ^{v}a^{\dag }-(a^{\dag }a\rho ^{v}+\rho ^{v}a^{\dag }a)/2$.

\begin{center}
\textbf{3. Numerical treatment and current formula}
\end{center}

In this section, we give a brief introduction to our mathematical approach. We pay our attention to the character of stationary transport. Therefore, the electron transmission can be conveniently described by the rate of electrons collected in the right lead. The system current is calculated according to the formula~\cite{Gurvitz:2005}
\begin{eqnarray}
I=-e\sum_{v=0}^{\infty }v\dot{P}^{v}.
\end{eqnarray}
Here, $P^{v}=Tr_{mech}[Tr_{char}[\rho ^{v}]]$ is the probability of $v$ electrons arrived at the right lead. The trace $Tr_{mech}$ is taken over variables of the mechanical oscillator and $Tr_{char}$ is taken over the degree of freedom of electron occupation in the QDs. For the following numerical treatment, we consider the wide band approximation and apply energy independent transmission rates $\Gamma_{ijy}=2\pi N_{y}T_{iy}^{\ast }T_{jy}$ ($i,j=1,2$ and $y=l,r$) and the damping rate $\gamma =2\pi D|g|^{2}$. We assume $\Gamma _{12y}$ and $\Gamma _{21y}$ are real and satisfy $\Gamma_{12y}=\Gamma _{21y}=\sqrt{\Gamma _{11y}\Gamma _{22y}}$. The chemical potentials for the left and right electrodes are set to be $\mu_{l}=eV/2$ and $\mu_{r}=-eV/2$, respectively.

The Fock state will be applied for the representation of the master equation. And so the Hilbert space of the system is generated by the composite basis $\{|00\rangle ,|01\rangle,|10\rangle ,|11\rangle \} \otimes \{|0\rangle \rangle ,|1\rangle\rangle ,...|n\rangle \rangle ,...\}$, where $\otimes $ means direct product. The state $|ij\rangle $ represents $i$ electrons in QD1 and $j$ electrons in QD2. $|n\rangle \rangle $ is the eigenstate of the $n$th excited level of the mechanical oscillator. In the Hilbert space, the system density matrix elements are written as $\rho _{ijkl,mn}=\langle \langle m|\otimes\langle ji|\rho |kl\rangle \otimes|n\rangle \rangle$, where $i,j,k,l=0,1$ and $m,n=0,1,2,...$ For any two given vibrational states $|m\rangle \rangle, |n\rangle \rangle$ we have $16$ density matrix elements $\rho _{ijkl,mn}$($i,j,k,l=0,1$) in terms of the electronic states. However, just $6$ of them are enough to describe the transport process since they can constitute a closed equation set for the system dynamics. These matrix elements are $\rho _{ijij,mn}$ and $\rho _{jiij,mn}$ ($i,j=0,1$).

For the case of strong inter-dot Coulomb interaction, we assume that the state of two-electron occupation is not inside the transport window. In other words, the bias voltage is so low that only one electron passes the system at any time. As a consequence, the process involving the state $\rho _{1111,mn}$ is not contained in our equations~\cite{Gurvitz:1996}. Substituting equation (7) into equation (13), we reach the following expression for the current
\begin{eqnarray}
I=I_{1}+I_{2}+I_{12},
\end{eqnarray}
where
\begin{widetext}
\begin{eqnarray}
I_{1}=eTr_{mech}[\Sigma^{out}_{11r}(0,0,1)\rho_{1010}-\Sigma^{in}_{11r}(0,0,1)\rho_{0000}],
\end{eqnarray}
\begin{eqnarray}
I_{2}&=&\frac{e}{2}e^{\alpha^{2}}\sum_{m_{1},n_{1},m_{2},n_{2}=0}^{\infty}
\frac{(\alpha)^{m_{1}+n_{1}+m_{2}+n_{2}}}{m_{1}!n_{1}!m_{2}!n_{2}!} \notag \\
&&\times Tr_{mech}[\Sigma^{out}_{22r}(m_{1},n_{1},2)((a^{\dag})^{m_{1}}(a)^{n_{1}}(a^{\dag})^{m_{2}}(a)^{n_{2}}
+(a^{\dag})^{m_{2}}(a)^{n_{2}}(a^{\dag})^{n_{1}}(a)^{m_{1}})\rho_{0101} \notag \\
&&-\Sigma^{in}_{22r}(m_{1},n_{1},2)((a^{\dag})^{n_{1}}(a)^{m_{1}}(a^{\dag})^{m_{2}}(a)^{n_{2}}
+(a^{\dag})^{m_{2}}(a)^{n_{2}}(a^{\dag})^{m_{1}}(a)^{n_{1}})\rho_{0000}],
\end{eqnarray}
\begin{eqnarray}
I_{12}=\frac{e}{2}e^{\alpha^{2}/2}\sum_{m,n=0}^{\infty}\frac{(\alpha)^{m+n}}{m!n!} Tr_{mech}[e^{-i\phi/2}(\Sigma^{out}_{12r}(0,0,1)(a^{\dag})^{m}(a)^{n}
+\Sigma^{out}_{12r}(m,n,2)(a^{\dag})^{n}(a)^{m})\rho_{0110}+h.c.].
\end{eqnarray}
\end{widetext}
Here, $I_{1}$ is the current through QD1 alone, $I_{2}$ is the current across the electromechanical junction in the absence of the reference arm. In fact, it is the same as the current directly derived from the master equation of the single-molecular junction~\cite{Lai}. And $I_{12}$ is the interference part in terms of the off-diagonal density matrix for the electronic states. The values of density matrix elements are achieved by solving equation (7) under the condition $\dot{\rho}=0$. We project the equation in the basis of the Hilbert space as
\begin{eqnarray}
\langle \langle m|\langle j i|(\varrho _{0}+\varrho _{1}+\varrho_{2}+\varrho _{12}+\varrho _{d})|i j\rangle|n\rangle \rangle=0,
\end{eqnarray}
\begin{eqnarray}
\langle \langle m|\langle i j|(\varrho _{0}+\varrho _{1}+\varrho_{2}+\varrho _{12}+\varrho _{d})|i j\rangle|n\rangle \rangle=0,
\end{eqnarray}
where $i,j=0,1$, excluding the case $i=j=1$. Then, one obtains a set of $5(N+1)^{2}$ linear equations. $N=0,1,2,...$ is the number of excited vibrational levels considered here. These equations can be solved by associating with the normalization condition $\sum_{v=0}^{\infty }P^{v}=1$. For the numerical treatment, we take $N=18$. It is a good approximation in the regime of low bias voltage, weak dot-lead couplings and finite oscillator damping rate as applied here, since the contribution from the higher levels ($n>N$) of the vibrational mode is very small.

\begin{center}
\textbf{4. Results and discussions}
\end{center}

\subsection{Phase relaxation and visibility}

\begin{figure}[!htb]
\centering \includegraphics[width=8cm]{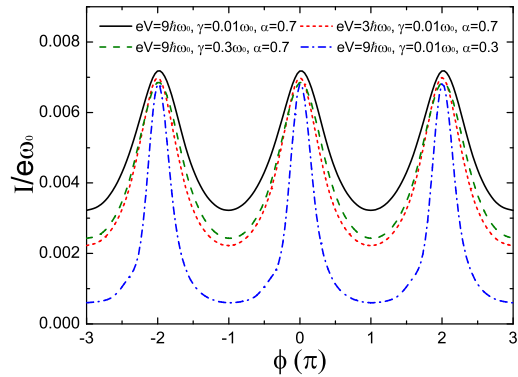}
\caption{(Color on line) Current as a function of the magnetic flux through the AB ring. For the solid black line we take the transmission rates $\Gamma_{22l}=\Gamma_{22r}=0.001312\omega_{0}$. For the red dotted line corresponding rates are $\Gamma_{22l}=\Gamma_{22r}=0.002731\omega_{0}$. For the green dashed line they are $\Gamma_{22l}=\Gamma_{22r}=0.003674\omega_{0}$ and for the blue dot-dashed line the transmission rates are $\Gamma_{22l}=\Gamma_{22r}=0.008091\Gamma$. The rest parameters are the same for all of the curves as $\varepsilon_{1}=\varepsilon_{2}=0$, $\Gamma_{11l}=\Gamma_{11r}=0.01\omega_{0}$, $k_{B}T=0.03\hbar\omega_{0}$, $x_{0}/d=0.003$.}
\label{iphi}
\end{figure}

In interference of two waves, visibility can be reduced not only by phase destruction of the waves but also due to difference of the absolute values of their amplitudes. The electromechanical systems substantially enhance electron transport for certain bias voltage~\cite{Park,Moskalenko,Boese,McCarthy}. Therefore, to see the net contribution of phase relaxation to the interference fringe, we balance the amplitudes of waves in the two paths by taking the bare transmission rates of QD2 smaller than that of QD1. To this end, the bare tunneling rates for the reference path are set to be $\Gamma _{11l}=\Gamma _{11r}=0.01\omega_{0}$ and for the target path are taken as $\Gamma_{22l}=\Gamma_{22r}=0.001312\omega_{0}$. Then, the current in equation (15) nearly equals to that in equation (16) with the small difference $I_{1}-I_{2}<10^{-5}/e\omega_{0}$. In this case, we suppose that the absolute values of the two amplitudes corresponding to the two paths are almost the same. With the above parameters, current versus magnetic flux is plotted by the solid line in figure~\ref{iphi}. It shows AB oscillation with the period of $2\pi$. Obviously, the weakest values of current at the points ($(2n+1)\pi$, $n$ is integer) of magnetic flux are not destructively interfered to be zero. It reveals that coherence of electron wave is influenced by the electromechanical vibration. Using the same way of balancing current amplitudes in the two paths, we give other three examples in figure~\ref{iphi} for different parameters. The low bias voltage $eV=3\hbar\omega_{0}$ (red dotted line), high damping rate $\gamma=0.3\omega_{0}$ (green dashed line) and small tunneling length $\alpha=x_{0}/\lambda=0.3$ (blue dot-dashed line) weaken the effect from vibrational mode. As a result, the interference fluctuation is enhanced.

\begin{figure}[!htb]
\centering \includegraphics[width=9cm]{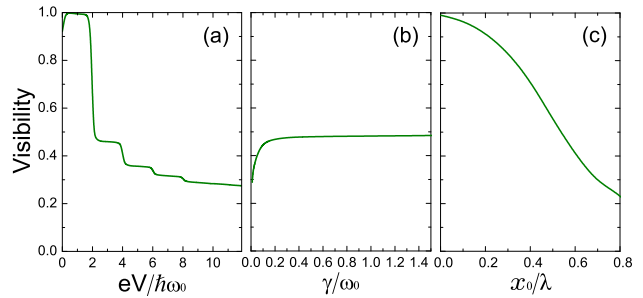}
\caption{(Color on line) (a) The visibility versus the bias voltage ($\varepsilon_{1}=\varepsilon_{2}=0$, $\Gamma_{11l}=\Gamma_{11r}=\Gamma_{22l}=\Gamma_{22r}=0.01\omega_{0}$, $\alpha=0.7$, $\gamma=0.01\omega_{0}$, $k_{B}T=0.03\hbar \omega_{0}$ and $x_{0}/d=0.003$). (b) The visibility as a function of the oscillator damping rate ($eV=9\hbar\omega_{0}$, $\varepsilon_{1}=\varepsilon_{2}=0$, $\Gamma_{11l}=\Gamma_{11r}=\Gamma_{22l}=\Gamma_{22r}=0.01\omega_{0}$, $\alpha=0.7$, $k_{B}T=0.03\hbar \omega_{0}$ and $x_{0}/d=0.003$). (c) The visibility as a function of the tunneling length ($eV=9\hbar\omega_{0}$, $\varepsilon_{1}=\varepsilon_{2}=0$, $\Gamma_{11l}=\Gamma_{11r}=\Gamma_{22l}=\Gamma_{22r}=0.01\omega_{0}$, $\gamma=0.01\omega_{0}$, $k_{B}T=0.03\hbar\omega_{0}$ and $x_{0}/d=0.003$).}
\label{visibi}
\end{figure}

Figure~\ref{iphi} shows the minimum and maximum values of the interference pattern is not shifted remarkably under different parameter variance. Using this feature, current visibility can be calculated easily by making a replacement as $I_{max}\simeq I(\phi =0)$ and $I_{min}\simeq I(\phi=\pi )$. It works under the conditions of $\varepsilon _{1}=\varepsilon _{2}=0$ and $\Omega\ll\omega_{0}$. The visibility of interference fringe is given by the formula $Visibility=(I_{max}-I_{min})/(I_{max}+I_{min})$. In figure~\ref{visibi} (a), a substantial influence of the bias voltage to the interference visibility can be seen. For very low voltages $eV<2\hbar\omega_{0}$, there is no exited level contained in the transport window ($eV$), and visibility is close to unity. When applied voltage is close to zero, the corresponding current approaches to be vanished. It causes a small drop of visibility near the zero voltage. Excited states of the mechanical oscillator play an important role for the phase relaxation of electrons. Increasing the bias voltage, excited levels of the vibrational mode are involved in the transport, which suppresses the visibility. In low voltage area, a few discrete states of the vibration contribute to the transport and the visibility displays a step profile. The mechanical oscillation is naturally coupled to the thermal bath and it has an intrinsic life time which is the inverse of the damping rate $\gamma$. By rising the damping rate, the visibility increase can be observed as shown in figure~\ref{visibi} (b). It is not hard to be understood that contribution from the mechanical motion would be decreased in the case of a high damping rate. The visibility is no longer enhanced obviously for damping rate $\gamma>0.2\omega_{0}$ . It is in accord with the transition of the electromechanical system from so called shuttling regime into tunneling regime~\cite{Novotny:2003,Novotny:2004}. The visibility is still not very high even at the quality factor $\omega_{0}/\gamma<1$. It implies, for the large damping rate, that coherence of electron does not obviously depend on the intrinsic lift time of the mechanical oscillator. In fact, the interference pattern is essentially affected by the strength of electron-phonon interaction which is determined by the parameter $\alpha=x_{0}/\lambda$. In figure~\ref{visibi} (c), the visibility versus the coupling strength $\alpha$ is plotted. For a given oscillator with mass $m$ and frequency $\omega_{0}$, the zero point uncertainty $x_{0}$ is fixed, and the coupling strength is mainly related to the tunneling length $\lambda$. For infinite large tunneling length $\lambda\rightarrow\infty$ we have $\alpha\rightarrow0$. In this case, $\varrho _{2}$ in equation (7) is close to the form of $\varrho _{1}$ and the effect of vibration in $\varrho _{2}$ and $\varrho _{12}$ approaches to be disappeared. As a consequence, tunneling between the two electrodes and QD2 is almost independent of the dot displacement. Therefore, we obtain that the visibility closes to be unity as illustrated in figure~\ref{visibi} (c).

In general, a large current induced by the vibrational junction is a reason of the reduction in visibility in the AB ring. For instance, when one takes the same bare tunneling rates for the two paths as shown in figure~\ref{visibi}, the probability of an electron passing the target arm is much larger than that of the electron propagating through the reference arm. There is another probable reason for the weak interference, namely phase shift of electron waves, and this is also our main interest in the present paper. The propagating of an electron wave through the vibrational junction gives rise to many scattering excited states. Especially at a sufficiently high applied voltage, the electron is in superposition of a large number of single-particle excited modes associated with the electron-phonon interaction. These excited states are characterized by phase acquirement related to absorbtion and emission of phonons. One can expect that a scattering in the space of positive phase shifts symmetrically happens in the space of negative phase shifts. As a consequence, interference of all the scattering waves does not exhibit global phase shift between the two paths. This property is valid for the same dot levels, $\varepsilon _{1}=\varepsilon _{2}$ and the weak charge-field coupling $\hbar \Omega $. In the next subsection let us discuss the case $\varepsilon _{1}\neq\varepsilon _{2}$. The influence of the charge-field coupling to the electron coherence has been previously considered in a similar system~\cite{Armour:2001,Armour:2002}.

\subsection{Coherent phase shift}

\begin{figure}[!htb]
\centering \includegraphics[width=8cm]{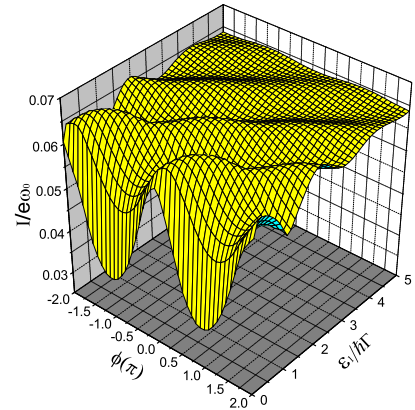}
\caption{(Color on line) AB interference oscillation as a function of the resonant level of QD1 ($eV=11\hbar\omega_{0}$, $\varepsilon_{2}=0$, $\Gamma_{11l}=\Gamma_{11r}=\Gamma_{22l}=\Gamma_{22r}=0.01\omega_{0}$, $\alpha=0.7$, $\gamma=0.01\omega_{0}$, $k_{B}T=0.03\hbar \omega_{0}$ and $x_{0}/d=0.003$)}
\label{iphieps3d}
\end{figure}

As we mentioned above, there is no global phase shift when an election is propagating through the single-dot electromechanical system. But it does not mean there is no phase shift when one component of the electron wave transports through any individual level of the system. In order to observe the phase change of propagating electron wave through the target system, we change gate voltage in the reference arm and see the variance of AB interference oscillation. As illustrated in figure~\ref{iphieps3d}, the pattern of the interference oscillation is shifting continuously in one direction when the resonant level of the reference arm is moving. The phase shift breaks original symmetry of the interference fringe for the replacement of $\phi$ by $-\phi$. It is induced by detuning between the two QDs. In the AB interferometer of electron transport, interference is remarkably strong only when the energy level in one path is close to that in the other path~\cite{Dominguez}. In other words, propagating waves in the two path are required to be oscillating in (at least nearly) the same frequency. Although the QD1 is detuned from the electronic level of the molecular junction, the molecular system still has energy levels provided by the mechanical oscillator. Therefore, interference is not disappeared in the case of the detuning except some phase shift.

The propagating wave in the reference arm only interferes with the wave in the molecular junction whose resonant energy ($\varepsilon _{2}+\Delta \varepsilon$) is the same as the resonant energy ($\varepsilon _{1}$) of the reference arm. Here, $\Delta \varepsilon$ is defined as the energy acquired or lost by an electron due to its inelastic scattering on the vibrating QD. Therefore, in figure~\ref{iphieps3d}, the phase shift corresponding to the detuning $\varepsilon _{1}-\varepsilon _{2}$ represents phase change of the sub transmission amplitude whose energy is $\varepsilon _{1}=\varepsilon _{2}+\Delta \varepsilon$ in the molecular junction. The total amplitude of electronic wave transferring through the molecular junction is, of course, superposition of all the sub transmission amplitudes with different resonant energies.

\begin{figure}[!htb]
\centering \includegraphics[width=8cm]{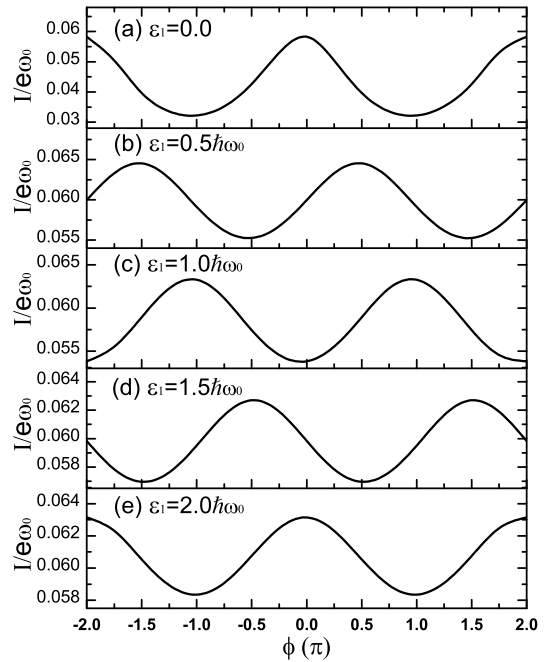}
\caption{Currents versus the variation of magnetic flux are plotted by changing the gate voltage of QD1 ($eV=9\hbar\omega_{0}$, $\varepsilon_{2}=0$, $\Gamma_{11l}=\Gamma_{11r}=\Gamma_{22l}=\Gamma_{22r}=0.01\omega_{0}$, $\alpha=0.7$, $\gamma=0.01\omega_{0}$, $k_{B}T=0.03\hbar \omega_{0}$ and $x_{0}/d=0.003$)}
\label{iphieps}
\end{figure}

By choosing particular detunings $\varepsilon _{1}-\varepsilon _{2}$ in figure~\ref{iphieps}, we analyze quantitative phase shift, especially for the resonant levels. Without loss of generality, the zero point energy is taken at $\varepsilon _{2}=0$. The numerical results in figure~\ref{iphieps} show the phase shift $\Delta\theta$ of the transmission amplitude with energy $\varepsilon _{2}+\Delta \varepsilon$ in the molecular junction roughly satisfies the relation $(\varepsilon_{1}-\varepsilon_{2})/\hbar\omega_{0} \simeq \Delta\theta/\pi$. When $\Delta \varepsilon=\Delta n\hbar \omega _{0}$ ($\Delta n=0,1,2,...$), the relation becomes $\Delta n \simeq \Delta\theta/\pi$. $\Delta n$ is defined as the net number of phonons involved in the inelastic transport. It intimates that the phase difference of propagating waves corresponding to two adjacent vibrational levels is $\pi$. This off-phase character is analogous to the phenomenon described by the Friedel sum rule~\cite{Friedel}. The sum rule relates the phase shift of a scattering electron to the number of states in the energy interval due to the scattering. However, the electron number accumulated in the impurity which is described by the general Friedel sum rule is replaced by the phonon number involved in the electron transfer in our present system.

The phase shift is very sharp when the energy of incident electron sweeping over the resonant levels~\cite{Pastawskia}. However, in our model the tunneling depends on the position of mechanical oscillator, so the phase change is continuous and very smooth. It is investigated by the fact that the position dependent tunneling causes inelastic process, which improves decay of the oscillating QD and broadens the energy levels of the system~\cite{Ueda}.

From figure~\ref{iphi}, we know that the changes in the applied voltage, the electron-phonon coupling $\alpha$ and the damping rate of the oscillator do not induce global phase shift in the AB interferometer. Therefore, the definitive phase relation of $\pi$ difference between two adjacent levels is independent of these parameters so long as they are properly taken that the discrete levels of the mechanical vibration effectively contribute to the electron transport. For instance, on one hand the applied voltage should be large enough that at least one excited level of the oscillator is included in the transport window. On the other hand, the voltage is not too large so that the feature of discrete levels involved in the tunneling is obviously manifested. In fact, the phase shift is just related to the unit quanta of the mechanical oscillator as shown in figure~\ref{iphieps}.

According to the above analysis, the neighboring resonant levels in the molecular vibrational junction are off-phase by $\pi$. It is the character of one dimensional quantum system which is considered in our model. Since, in one dimensional system, the upper energy level has $1$ more wave function node than the lower one, and each node changes the phase of transmission amplitude by $\pi$. This property may be not true if the system is not strictly one dimensional~\cite{Lee}. In the experiment of AB interferometer where a fixed QD is embedded in one of the arms, the phase behaviors are the same for all resonant levels of the QD~\cite{Yacoby,Schuster,Cernicchiaro}. Namely, all the resonant levels are in-phase. It is different from present effect found in the electromechanical system, where all the vibrational levels are coherently correlated with definitive phase difference of $\pi$. The phase shift varies from $0$ to any large value, depending on the net number of phonons involved in an electron tunneling.

The reason of the visibility depression mentioned in the last subsection becomes more clear now. Actually, an electron takes all the channels of the discrete vibrational levels which are involved in the transport process. Therefore, interference not only occurs between the propagating waves in the two paths, but also occurs among the waves taking different channels of the vibrational junction. As we analyzed above, any two neighboring channels have a phase difference of $\pi$. The wave functions taking different vibrational levels destructively interfere because of the phase differences. It is the reason of phase relaxation in the AB interference due to the vibrational junction (see figure~\ref{iphi}). It has been shown in a double-QD two-electron AB interference that two components of conductance oscillations with the same amplitudes cancel each other due to their phase difference of $\pi$~\cite{Akera}. Since two components of the conductance are the same in amplitude, the final conductance disappears in their system. In the present case, the electron occupation probabilities on different energy levels of the electromechanical system are not the same. Therefore, there is net current remained in the system, but it is not fully coherent. In fact, the interference between the different channels is also reflected in the direct transmission of charge through the electromechanical system~\cite{Lai}. The current calculated from the scheme considering both diagonal and off-diagonal density matrix elements of the system is remarkably lower than that obtained by the approach in which only diagonal terms are taken into account. This current suppression is related to the destructive interference between different transport channels.

\begin{center}
\textbf{5. Conclusions}
\end{center}

Electrons propagating through the single-molecular vibrational junction are dephased. It is caused by the electron scattering on the excited levels of the vibrational mode. However, the interference fringe of the AB interferometer is not absolutely destroyed by the nanoelectromechanical system. The visibility is sensitive to the applied voltage, the oscillator damping rate and the tunneling length. The transmission amplitudes corresponding to channels of the vibrational resonant levels are coherently correlated via any neighboring channels have a definitive phase difference of $\pi$. Because of the phase shifts between the resonant levels in the electromechanical junction, different branches of the transmission waves destructively interfere with each other. As a consequence, the electron tunneling through the system appears not to be fully coherent. The character of the phase difference of $\pi$ is robust with respect to wide range of the bias voltage, the tunneling length and the life time of the vibrational mode. It just depends on the frequency of the mechanical oscillator. This work would provide a guidance for the experimental observation of dephasing in electron transport through a vibrational molecular junction and phonon assisted conductors.

\begin{acknowledgments}
We acknowledge the supports from NNSFC Grant (91021017, 11274013) and NBRP of China (2012CB921300).
\end{acknowledgments}

\end{document}